Atmospheric Science Letters — RMetS

RESEARCH ARTICLE

# Identification of synoptic weather types over Taiwan area with multiple classifiers

Shih-Hao Su[1] | Jung-Lien Chu[2] | Ting-Shuo Yo[3,4] | Lee-Yaw Lin[2]

[1]Chinese Culture University, Taipei, Taiwan

[2]National Science and Technology Center for Disaster Reduction, Taipei, Taiwan

[3]National Taiwan University, Taipei, Taiwan

[4]DataQualia Lab Co. Ltd., Taipei, Taiwan

**Correspondence**
Jung-Lien Chu, National Science and Technology Center for Disaster Reduction, Taipei, Taiwan.
Email: jlchu@ncdr.nat.gov.tw

**Funding information**
Ministry of Science and Technology, Taiwan, Grant/Award Number: MOST-104-2111-M-034-003, MOST-104-2625-M-034-002, MOST-105-2111-M-034-003, MOST-106-2621-M-865-001

In this study, a novel machine learning approach was used to classify three types of synoptic weather events in Taiwan area from 2001 to 2010. We used reanalysis data with three machine learning algorithms to recognize weather systems and evaluated their performance. Overall, the classifiers successfully identified 52–83% of weather events (hit rate), which is higher than the performance of traditional objective methods.

The results showed that the machine learning approach gave low false alarm rate in general, while the support vector machine (SVM) with more principal components of reanalysis data had higher hit rate on all tested weather events. The sensitivity tests of grid data resolution indicated that the differences between the high- and low-resolution datasets are limited, which implied that the proposed method can achieve reasonable performance in weather forecasting with minimal resources.

By identifying daily weather systems in historical reanalysis data, this method can be used to study long-term weather changes, to monitor climatological-scale variations, and to provide better estimate of climate projections. Furthermore, this method can also serve as an alternative of model output statistics and potentially be used for synoptic weather forecasting.

**KEYWORDS**

front, heavy rainfall, machine learning, synoptic weather classification, Taiwan, typhoon

## 1 | INTRODUCTION

Most extreme weather events in Taiwan are associated with strong synoptic-scale systems. The fluctuations of these weather systems reflect variations in climatological scale and induce a variety of mesoscale mechanisms. These results in precipitation and temperature change with different temporal and spatial characteristics, and hence are the major cause of natural disasters. Therefore, the recognition of various types of synoptic weather is crucial for weather forecasting, disaster prevention, and climate projections.

Before 1960, the common weather classification techniques used by most meteorologists were mainly manual and subjective methods (e.g., Lamb, 1950). Since early 1960s, objective diagnosis methods were developed with advances in computer technologies and grid data. Renard and Clarke (1965) used spatial gradients of atmospheric thermal and moment parameters to identify weather systems. Their study suggested that using simple thermal and dynamic parameters alone was not robust for the identification of weather fronts.

Meanwhile, other researchers proposed statistical and similarity-based methods for weather typing. Some studies used exemplars of weather systems and similarity metrics, such as spatial correlation (Lund, 1963) or sum of squared difference (Kirchhofer, 1973), to identify their occurrences. Other methods used exploratory analysis techniques to identify the major spatial patterns of meteorological data and associated them with weather systems (Richman, 1981; Key and Crane, 1986; Huth, 1996a; 1996b).

As machine learning gained its popularity in scientific research, techniques such as self-organizing maps was used for weather typing and classifications (Jiang *et al.*, 2012; Hope *et al.*, 2014). In 2016, Liu and colleagues applied convolutional deep neural networks to detect extreme weather systems in simulated and reanalysis datasets (Liu, 2016). Their results showed that machine learning techniques were suitable for detecting weather systems.

In present study, we proposed a classification-based approach with long-term reanalysis data. We introduced the grid numerical outputs and observational records for model training and used machine learning techniques to classify the synoptic weather types from 2001 to 2010. The climate reanalysis data and surface observational records are described in next section. The automatic analysis methods and the results are presented in sections 3 and 4. Section 5 discusses the results and the usage of the automatic weather classifier.

## 2 | DATA SOURCES

In this study, we focused on three weather events, namely fronts, typhoons, and heavy rainfall (HR) events. These events are three major weather types associated with nature disasters in Taiwan. The records of these events were sourced from a newly developed dataset referred to as the Taiwan Atmospheric Events Database (TAD, personal connection). This dataset consists of major synoptic-scale weather events in Taiwan area identified with objective and subjective methods. The front events during 2001–2010 were identified with subjective surface analysis of the Central Weather Bureau (CWB) weather maps. The selected domain was used to identify the fronts near Taiwan, in the area 119°–123°E and 21°–26°N. Typhoon events were determined by the hourly typhoon center position data according to the CWB typhoon database (Wang, 1980). The CWB hourly precipitation data from 31 manual observation stations and more than 690 automatic rain gauges were used to identify HR events in Taiwan. The criteria of HR, that is, 80 mm/day or 40 mm/hr, was enforced by the CWB of Taiwan.

The National Centers for Environmental Prediction (NCEP) Climate Forecast System Reanalysis (CFSR; Saha *et al.*, 2010; Saha *et al.*, 2014) was used for training and evaluation of our statistical model in this study. The CFSR reanalysis data repository provided highly detailed temporal (6 hr) and spatial (0.5 × 0.5°) information from 1979 to 2010; these datasets are comparable to many other sets of reanalysis data. We also used coarse resolution global grid data from the European Centre for Medium-Range Weather Forecasts (ECWMF), ECMWF's atmospheric reanalysis of the 20th century (ERA-20C, 2.5 × 2.5° horizontal resolution; Poli *et al.*, 2016) to examine the numerical grid resolution effects. The variables used in this study are summarized in Table 1a.

## 3 | AUTOMATIC ANALYSIS TECHNIQUES

In this section, the architecture of the proposed weather classification approach is detailed. In addition to the method, a series of experiments was designed to evaluate the effectiveness of the system. We investigated the classification performance of several preprocessing settings, feature sets, and machine learning algorithms.

For a classification task, two types of data are required: input data and output. Input data consists of the feature sets upon which decisions are made. The output variables represent the target decisions or predictions. In the proposed procedure, the output is a set of binary labels indicating whether a weather event occurs.

In this study, we used two sources of data as inputs: coarse resolution global grid data from ERA-20C and fine numerical grid analysis fields from CFSR over the East Asia region. The choice of data sources was made to increase the variety of the feature space, so that the classification results can show which type of data is more informative for a certain type of weather event.

**TABLE 1** The list of (a) selected variables of grid data used in experiments and (b) feature sets used in the experiments

| (a) Input variable | | Symbol | Selected levels |
| --- | --- | --- | --- |
| Zonal and meridional wind components | | U, V | 925, 850, 700, and 200 hPa |
| Temperature | | T | 925, 850, 700, and 200 hPa |
| Dew point temperature | | Td | 925, 850, and 700 hPa |
| Mean sea level pressure | | MSLP | Surface |
| Geopotential height | | H | 500 hPa |
| (b) Input variable | Symbol | Operational definition | |
| High-resolution grid data | NCEP50 | The principle components that explain 50% of variance of each variable of NCEP-CFSR grid data | |
| | NCEP70 | The principle components that explain 70% of variance of each variable of NCEP-CFSR grid data | |
| Low-resolution grid data | EC50 | The principle components that explain 50% of variance of each variable of ECMWF ERA-20C grid data | |
| | EC70 | The principle components that explain 70% of variance of each variable of ECMWF ERA-20C grid data | |
| Multi-resolution grid data | EC-NCEP50 | The principle components that explain 50% of variance of each variable of both of CFSR and ERA-20C grid data | |
| | EC-NCEP70 | The principle components that explain 70% of variance of each variable of CFSR and ERA-20C grid data | |

The data were further filtered and processed as follows. For the reanalysis data from both ERA-20C and CFSR, 17 vertical levels of 00Z each day were selected. The selected variables consisted of the mean sea level pressure; U, V, Td, and T of 925, 850, and 700 hPa; geopotential height of 500 hPa; and T, U, and V of 200 hPa. The randomized Principal Components Analysis (PCA) was then applied to each variable for dimension reduction (Halko et al., 2011; Pedregosa et al., 2011). And thus, the first few principal components (PCs) that can explain K% of total variance of each variable were selected as the feature set. The number of PCs selected was decided based on the proportion of variance can be explained, and hence is different for each variable. Generally speaking, most of 17 selected variables can be represented by a few PCs except the humidity parameters and wind field at 200 hPa. Also, the lower-resolution-global-domain dataset required more PCs to explain the same amount of variance compared to the other dataset. We experimented with different $K$ values and tried to balance between the model accuracy and complexity, and finally decided to present the $K$ values of 50 and 70, where a limitation of at most 100 PCs can be used for one single variable was enforced.

Table 1b shows six different feature sets used in our experiments. These sets were designed to show what types of data gave the most information indicating the occurrence of a given weather event. The feature set with different grid resolutions helped us determine which resolution of model data is suitable for use in operation.

From the wide variety of algorithms capable of binary classification, three different classifiers were selected for this study, namely the logistic regression model (GLM; R Core Team, 2015), gradient boosting model (GBM; Greg Ridgeway, 2015), and support vector machine (SVM; Karatzoglou et al., 2004) with polynomial kernel. These three classifiers represent three different approaches to classification: a linear model, an ensemble tree-based model, and a nonlinear model. The GLM was chosen because of its simplicity and explainability, and its performance often serves as a baseline for other classifiers. In a thorough review of different classifiers by Fernández-Delgado and his colleagues (Fernández-Delgado et al., 2014), random forest and SVM were suggested as the two best classification algorithms for most real-world data. Hence, their R implementations, that is, GBM and SVM, were chosen for the designed experiments.

## 4 | RESULTS

Weather forecasts of binary events were conventionally verified with hit rate ($H$), false alarm rate ($F$), false alarm ratio (FAR), and critical success index (CSI; Jolliffe and Stephenson, 2012). In the context of anomaly detection, positive predictive values (PPV) and F-1 score are commonly used for evaluation. Table 2 showed a basic confusion matrix, and its elements are used to explain the measurements of performance as follows.

Hit rate ($H$) is also known as sensitivity. It measures the proportion of positives that are correctly identified as such. Its mathematical form can be written as

$$\text{Hit rate}(H) = \frac{\text{Number of ture positive}}{\text{Total of true positives}} = \frac{A}{A+C}.$$

False alarm rate ($F$) measures the proportion of false positives over all negative cases, and a high hit rate represents that most events are detected by the system. The formula of false alarm rate is

$$\text{False alarm rate}(F) = \frac{\text{Number of false positive}}{\text{Total of true negatives}} = \frac{B}{B+D}.$$

FAR measures the proportion of cases identified as positives that are wrong. The mathematical expression of FAR is

$$\text{FAR} = \frac{\text{Number of false positive}}{\text{Total of preditcted positives}} = \frac{B}{A+B}.$$

CSI measures the conditional probability of a hit given that the event was either forecast, or observed, or both. CSI is often used for evaluating rare event detection, and can be mathematically expressed as

$$\text{CSI} = \frac{\text{Number of ture positive}}{\text{Total of classified positives} \wedge \text{true positives}} = \frac{A}{A+B+C}.$$

PPV refers the proportion of positive classified results that are true positive results, and it equals to $1 - \text{FAR}$. A high PPV means an event is more likely to occur when the system detects so. The equation for PPV is given as

$$\text{PPV} = \frac{\text{Number of ture positive}}{\text{Total of classified positives}} = \frac{A}{A+B}.$$

$F1$ score is the harmonic mean of hit rate and PPV, and it is commonly used in the field of signal processing for anomaly detection. The $F1$ score is similar to CSI except it gives higher weight to true positive cases. The mathematical form of $F1$ score can be written as

$$F1 = \frac{2*H*\text{PPV}}{H+\text{PPV}} = \frac{2A}{2A+B+C}.$$

**TABLE 2** The basic confusion matrix of weather classified experiments

| | Condition positive | Condition negative |
|---|---|---|
| Classified positive | True positive (hit) A | False positive (false alarm) B |
| Classified negative | False negative (miss) C | Ture negative (correct rejection) D |

The gray shaded cells are the bad classifications and white shaded cells are the good classifications.

All six measures described above, namely the *H*, *F*, FAR, CSI, PPV, and F-1, are calculated for each combination of event, feature set, and classifiers.

Besides the performance measures, the sampling scheme used in the experiments also affects the evaluation. Cross-validation (Stone, 1974) is a commonly used re-sampling technique known to provide a good estimate of the true out-of-sample performance. In this study, each dataset–classifier combination was evaluated with a 10-fold cross-validation scheme sampled with the same random seed.

Figure 1 depicts the results from the classification experiment. The proposed approach with SVM is shown to identify the typhoon, front, and HR events with hit rate of 37–64%, 26–52%, and 78–83%, and *F*1 score of 52–72%, 39–58%, and 77–82%, respectively. While hit rate indicates the ability that the system can identify an event when it occurs, *F*1 score balances the hit rate by penalizing false alarms. The results showed that the proposed method gave good hit rate and kept the false alarm rate low at the same time.

Figure 2 shows the *F*1 scores over three types of events. As illustrated in the figure, SVM with the PCs explained 70% of variance of NCEP-CFSR data outperformed other feature combinations in most cases. For HR events, SVM with NCEP-CFSR and ECMWF data performed the best, though adding ECMWF data gave only minor improvement. The comparison among the performance of different classifiers showed that the SVM is a capable choice for such tasks.

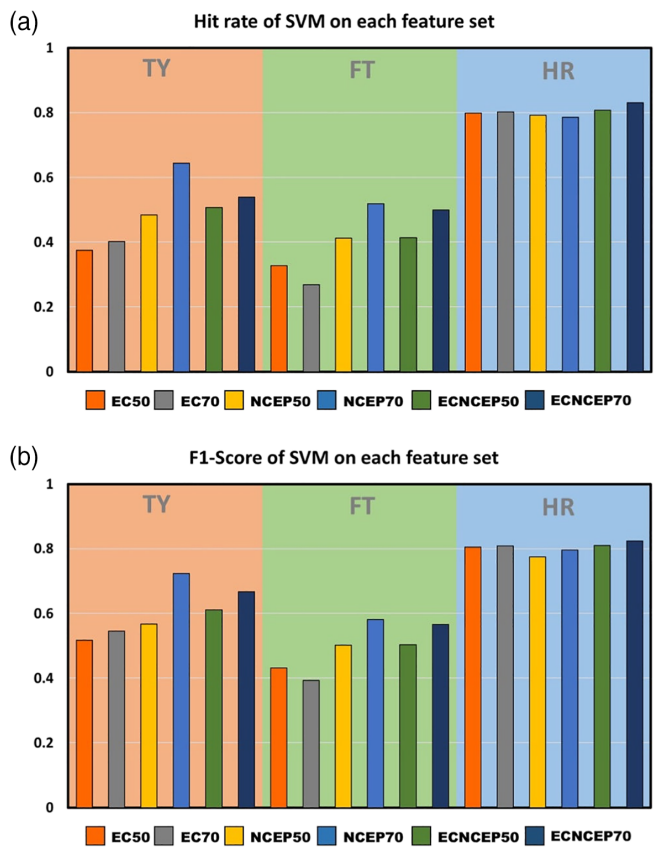

FIGURE 1  The (a) hit rate and (b) *F*1-score of SVM over three events

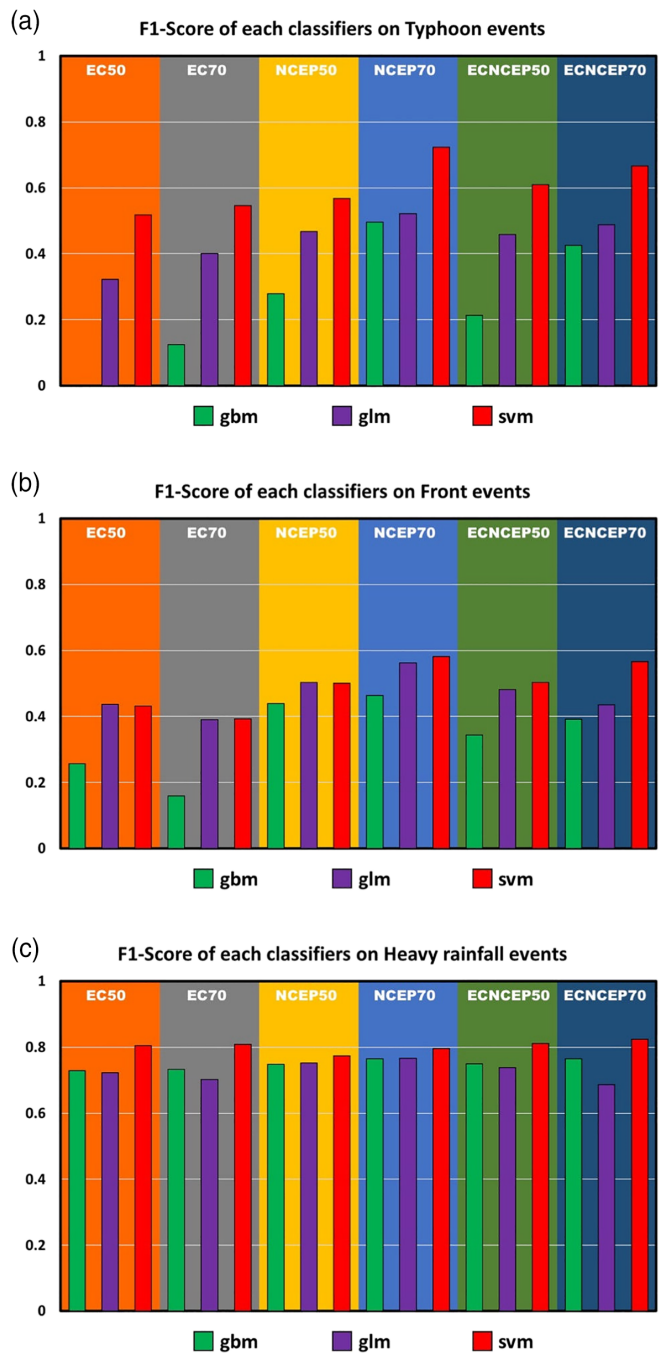

FIGURE 2  The *F*1-score of SVM on each feature set over three events

Results of different grid datasets, NCEP-CFSR and ECMWF, suggest that high-resolution regional data (0.5° over East Asia) are more informative than low-resolution global data (2.5° global). Although combining both datasets can provide more information, the improvement is minor or even negative, as shown in Figure 2.

The ROC curve of the results is shown in Figure 3. The figure shows that the proposed method tends to give higher hit rate in general. The overall performance (distance away the no-skill limitation) are much better than the traditional objective methods. Also, we noticed the performance of detecting HR events is more consistent than other events as illustrated in Figure 3.

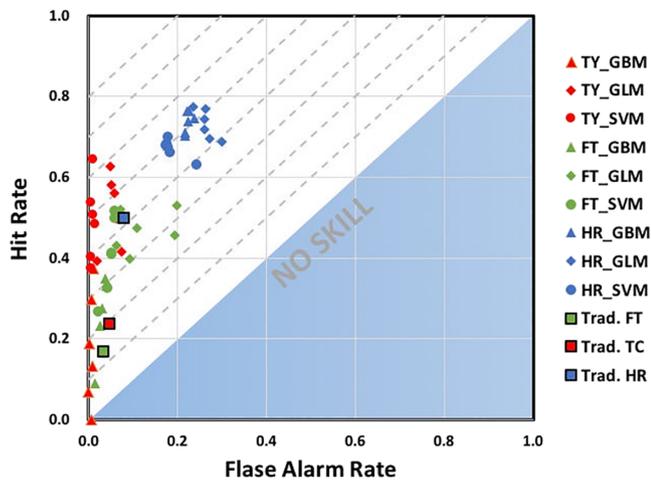

**FIGURE 3** The relative operating characteristic (ROC) diagram of all experiments. The shaded area marked the no-skill region and gray dash lines were the performance references. The red, green, and blue markers represented the typhoon, front, and HR events. Different symbols indicated three classifiers and black squared mark shows the traditional objective analysis

## 5 | DISCUSSION AND CONCLUSIONS

In this paper, we used the machine learning technique to classify three types of synoptic weather events in Taiwan area from 2001 to 2010. The results show better performance in comparison to objective analysis methods, and the details are discussed as follows.

For frontal system, Renard and Clarke (1965) first proposed using the "thermal front parameter (TFP)" to identify the system from grid data. They manually defined the front location based on TFP and reported a fair accuracy. Hope *et al.* (2014) used multiple TFP-like indices with artificial neural networks to recognize the Australian winter frontal system. They reported the equitable threat score (ETS) as 0.0–0.18 and the best ETS of in our study is 0.33 (SVM with fine-resolution whose PCA modes explained 70% of variance). We applied the TFP-based methods of Hope *et al.* (2014) to the NCEP-CFSR dataset, and the resulting hit rate is 0.17 with a false positive rate of 0.03. The hit rate and false positive rate of our method are 0.52 and 0.06, respectively.

Many methods were proposed to detect the tropical cyclone in the past studies. For comparison, we adopted an objective TC detection method proposed by Vitart *et al.* (1997) and applied it to the same NCEP-CFSR dataset. While the SVM showed a hit rate of 0.64, the objective TC detection method gave a hit rate of 0.24 from the same reanalysis inputs.

To compare the ability of recognizing HR events, NCEP-CFSR simulated precipitation rate was used as the baseline (Cheng *et al.*, 2011). Haiden *et al.* (2012) used similar model outputs to evaluate the Quantitative Precipitation Forecast (QPE) skill for multiple global models. According to their results, the QPF with model outputs showed excessive forecasting of light rain, and it had difficulty in predicting heavy rain events. In this study, the best ETS of HR classifiers is 0.48, which is higher than 0.003 directly from the QPF via NCEP-CFSR and 0.30–0.35 reported by Haiden *et al.* (2012) with resolution-independent thresholds.

In addition to the performance, the system achieved descent hit rate without high false alarm rate. The results demonstrated that the SVM with the PCs explained 70% of variance of NCEP-CFSR data gave better performance in general. The difference between NCEP-CFSR and ECMWF data can serve as sensitivity tests of grid data resolution. The corresponding results showed that high-resolution regional data is more informative than low resolution data for all events, though the difference was limited. Besides different number of principle components used, we also examined the additional input datasets (7–17 layers) of meteorological fields by selected strategies. The results showed that the classifiers improved by providing more information for most weather events except the frontal system. This may relate to the vertical structure difference of the seasonal frontal systems, suggesting further investigations of sub-types of the front events. In addition, according to our results, increasing the numerical model resolution only brings minor improvement. This suggests that one can obtain reasonable improvement in weather forecasting while adding minimal resources.

In this study, we have successfully demonstrated the use of machine learning methods for synoptic weather classification. The results showed that our method outperformed methods based on traditional objective diagnosis. The proposed method is equivalent to a pattern recognizer that identifies weather events from given reanalysis data, and it has many potential applications. For example, one may apply it to the historical reanalysis datasets and the results can be used to study long-term historical weather changes. This can help to monitor climatological-scale variations as well as to provide better estimation of climate projections. Furthermore, this method can also serve as an alternative of model output statistics (MOS) and potentially be used for synoptic weather forecasting.

### ACKNOWLEDGMENTS

We thank Dr. Yi-Liang Chen for his helpful comments and suggestions. We thank Bing-Kui Chiou and Kao-Yuan Liu for their help in processing the data. This research was supported by the grants from Ministry of Science and Technology of Taiwan through grants MOST-104-2111-M-034-003, MOST-104-2625-M-034-002, MOST-105-2111-M-034-003, MOST-106-2621-M-865-001 and to Chinese Culture University. This study was also collaborating with the Taiwan Climate Change Projection and Information Platform Project (TCCIP). We thank the Central Weather Bureau for their typhoon database and provided the rainfall data.

**ORCID**

*Jung-Lien Chu* http://orcid.org/0000-0002-8697-5831